%% file: typeinst.tex
\documentclass{llncs}

\usepackage{amssymb}
\usepackage{graphicx}

\usepackage{url}
\urldef{\mailsa}\path|{binxuanh, kathleen.carley}@cs.cmu.edu|

\usepackage{color}

\begin{document}

\mainmatter  

\title{On Predicting Geolocation of Tweets using Convolutional Neural Network}


%
%
\author{Binxuan Huang%
\and Kathleen M. Carley}
\authorrunning{Lecture Notes in Computer Science: Authors' Instructions}

\institute{School of Computer Science, Carnegie Mellon University,\\
5000 Forbe Ave., Pittsburgh, United States\\
\mailsa\\
}

%
%

\toctitle{Lecture Notes in Computer Science}
\tocauthor{Authors' Instructions}
\maketitle

\vspace{-0.5cm}
\begin{abstract}
\input{./0_abstract}
\vspace{-0.2cm}
\end{abstract}

\input{./1_intro_v2}

\input{./2_related}

\input{./3_method}

\input{./4_data}

\input{./5_exp_v2}

\input{./6_conclusion}

\subsubsection*{Acknowledgments.}This work was supported in part by the Office of Naval Research (ONR) N000140811186, and the National Science Foundation(NSF) 00361150115291. The views and conclusions contained in this
document are those of the authors and should not be interpreted as representing the official policies, either expressed or implied, of the Office of Naval Research or  the National Science Foundation. We want to thank tutors in the Global Communication Center at Carnegie Mellon for their valuable advice.
     	        \vspace{-0.2cm}
\bibliographystyle{splncs03.bst}
\bibliography{ref}
\end{document}

%% file: 0_abstract.tex
In many Twitter studies, it is important to know where a tweet came from in order to use the tweet content to study regional user behavior. However, researchers using Twitter to understand user behavior often lack sufficient geo-tagged data.  Given the huge volume of Twitter data there is a need for accurate automated geolocating solutions. Herein, we present a new method to predict a Twitter user's location based on the information in a single tweet. We integrate text and user profile meta-data into a single model using a convolutional neural network. Our experiments demonstrate that our neural model substantially outperforms baseline methods, achieving 52.8\% accuracy and 92.1\% accuracy on city-level and country-level prediction respectively.


%% file: 1_intro_v2.tex
\section{Introduction}
\vspace{-0.2cm}
Recently, there is growing interest in using social media to understand social phenomena. For example, researchers have shown that analyzing social media reveals important geospatial patterns for keywords related to presidential elections\cite{tsou2013mapping}. People can use Twitter as a sensor to detect earthquakes in real-time\cite{sakaki2010earthquake}. Recent research also has demonstrated that Twitter data provides real-time assessments of flu activity\cite{achrekar2011predicting}.

Using Twitter's API\footnote{https://dev.twitter.com/docs}, a keyword search can be done and we can easily get tweet streams from across the world containing keywords of interest. However, we cannot conduct a fine-grained analysis in a specific region using such a keyword-based search method. Alternatively, using the same API tweets with geo-information can be collected via a bounding box. Since less than 1\% of tweets are tagged with geo-coordinates\cite{hale2012world}, using this location-based search means we will lose the majority of the data. If we can correctly locate those ungeotagged tweets returned from a keyword search stream, that would enable us to study users in a specific region with far more information.

With this motivation, we are aiming to study the problem of inferring a tweet's location. Specifically, we are trying to predict on a tweet by tweet basis, which country and which city it comes from. Most of the previous studies rely on rich user information(tweeting history and/or social ties), which is time-consuming to collect because of the Twitter API's speed limit. Thus those methods could not be directly applied to Twitter streams. In this paper, we study a global location prediction system working on each single tweet. One data sample is one tweet JSON object returned by Twitter's streaming API. Our system utilizes location-related features in a tweet, such as text and user profile meta-data. We summarize useful features that can provide information for location prediction in {Table \ref{featuretable}}. 
\vspace{-0.5cm}
\begin{table}[!h]
\label{featuretable}
\centering
\caption{Feature table}
\begin{tabular}{|l|l|}
\hline
Feature                   & Type        \\ \hline
Tweet content             & Free text   \\ \hline
User personal description & Free text   \\ \hline
User name                 & Free text   \\ \hline
User profile location     & Free text   \\ \hline
Tweet language(TL)           & Categorical \\ \hline
User language(UL)             & Categorical \\ \hline
Timezone(TZ)                  & Categorical \\ \hline
Posting time(PT)		&	UTC timestamp \\ \hline
\end{tabular}
\end{table}
\vspace{-0.5cm}

Recent research has shown that using bag-of-words and classical machine learning algorithms such as Naive Bayes can provide us a text-based location classifier with good accuracy\cite{han2014text}. Different from previous research, we intend to use the convolutional neural network(CNN) to boost prediction power. Inspired by the success of convolutional neural network in text classification\cite{kim2014convolutional}, we are going to use CNN to extract location related features from texts and train a classifier that combines high-level text feature representations with these categorical features. To benchmark our method, we compared our approach with a stacking-based method. Experimental results demonstrate that our approach achieves 92.1\% accuracy on country-level prediction and 52.8\% accuracy on city-level prediction, which greatly outperforms our baseline methods on both tasks.

%% file: 2_related.tex
\section{Related work}
\vspace{-0.2cm}
Identifying demographic details of Twitter users\cite{mislove2011understanding} has been widely studied in previous literature including inferring users' attributes like age, gender\cite{culotta2015predicting}, political affiliation\cite{pennacchiotti2011democrats} and personalities like openness and conscientiousness\cite{quercia2011our}. Among these research, there is increasing interest in inferring Twitter user's location, which is largely driven by the lack of sufficient geo-tagged data\cite{hale2012world}. In many situations, it is important to know where a tweet came from in order to use the information in the tweet to effect a good social outcome. Key examples include: disaster relief\cite{landwehr2014social}, disaster management\cite{carley2016crowd}, earthquake detection\cite{earle2012twitter}, predicting elections\cite{shi2012predicting}, and predicting flu trends\cite{achrekar2011predicting}.

A majority of previous works either focus on a local region e.g. United States\cite{cheng2010you}, Sweden\cite{berggren2016inferring}, or using rich user information like a certain number of tweets for each user\cite{cheng2010you}, user's social relationship\cite{jurgens2013s,qian2017probabilistic,wang2014location}. Different from these works, this paper works on worldwide tweet location prediction. We only utilized features in one single tweet without any external information. Thus this method could be easily applied to real-time Twitter stream.

For fine-grained location prediction, there are several types of location representation methods existing in literature. One typical method is to divide earth into small grids and try to predict which cell one tweet comes from. Wing and Baldridge introduced a grid-based representation with fixed latitude and longitude\cite{wing2011simple}. Based on the similarity measured by Kullback-Leibler divergence, they assign each ungeotagged tweet to the cell with most similar tweets. Because cells in urban area tend to contain far more tweets than the ones in rural areas, the target classes are rather imbalanced\cite{han2013stacking}. To overcome this, Roller et al. further proposed an adaptive grid representation using K-D tree partition\cite{roller2012supervised}. Another type of representation is topic region. Hong et al. proposed a topic model to discover the latent topic words for different regions\cite{hong2012discovering}. Such parametric generative model requires a fixed number of regions. However, the granularity of topic regions is hard to control and will potentially vary over time\cite{han2014text}.

The representation we choose is city-based representation considering most tweets come from urban area. One early work proposed by Cheng et al. using a probabilistic framework to estimate Twitter user's city-level location based on the content of tweets\cite{cheng2010you}. Their framework tries to identify local words with probability distribution smoothing. However, such method needs a certain number of tweets(100) for each user to get a good estimation. Han et al. proposed a stacking-based approach to predict user's city\cite{han2013stacking}. They combine tweet text and meta-data in user profile with stacking\cite{wolpert1992stacked}. Specifically, they train a multinomial naive Bayes base classifier on tweet text, profile location, timezone. Then they train a meta-classifier over the base classifiers. More recently, Han et al. further did extensive experiments to show that using feature selection method, such as information gain ratio\cite{quinlan2014c4} and $\chi^2$ statistic could greatly improve the classification performance. 


%% file: 3_method.tex
\section{Location prediction}
\vspace{-0.2cm}
In this section, we will introduce our location prediction approach. We first briefly describe the useful features in a tweet JSON object. After that, we will further explain how we utilize these features in our prediction model.
\vspace{-0.2cm}
\subsection{Feature Set}
\vspace{-0.1cm}
We have listed all useful information we want to utilize in Table \ref{featuretable}. Tweet content, user personal description, user name and profile location are four text fields that we will use. Twitter users often reveal their home location in their profile location and personal description. However, location indicating words are often mixed with informal tweet text(e.g. chitown for Chicago). It is unrealistic to use a gazetteer to find these words. In this work, we choose to apply CNN on these four text fields to extract high-level representations.

In addition to these four text fields. there are another three categorical features: tweet language, user language,and timezone. Tweet language is automatically determined by Twitter's language detection tool. User language and timezone are selected by the user in his/her profile. These three categorical features are particularly useful for distinguishing users at the country-level.

\begin{figure}[h]
    \centering
    \vspace{-0.2cm}
    \includegraphics[width=0.5\textwidth]{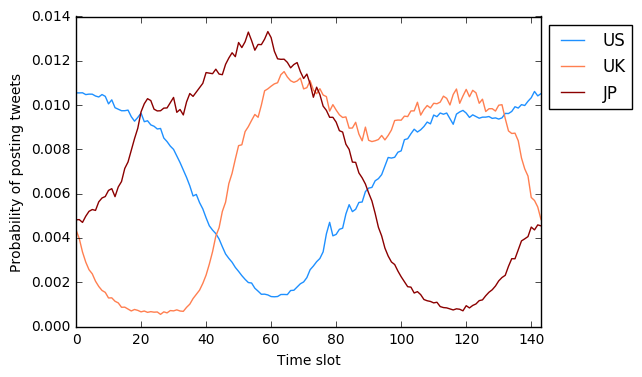}
    \vspace{-0.4cm}
    \caption{The probability of an user posting a tweet in different time slot in three different countries: United States, United Kingdom, Japan.}
    \label{time}
        \vspace{-0.6cm}
\end{figure}

The last feature is UTC posting time. Using posting timestamp as a discriminative feature is motivated by the fact that people in a region are more active on Twitter at certain times during the day. For example, while people in United Kingdom start to be active at 9:00 am in UTC time, most of the people in United States are still asleep. We transform the posting time in UTC timestamp into discrete time slots. Specifically, we divide 24 hours into 144 time slots each with a length of 10 minutes. Thus each tweet will have a discrete time slot number in the range of 144, which can be viewed as a categorical feature. In Figure \ref{time}, we plotted the probability distribution of an user posting tweets in each time slot in three different countries. As expected, there is a big variance between these three countries.
\vspace{-0.4cm}
\subsection{Our Approach}
\begin{figure}[h]
    \centering
    \vspace{-1cm}
    \includegraphics[width=0.8\textwidth]{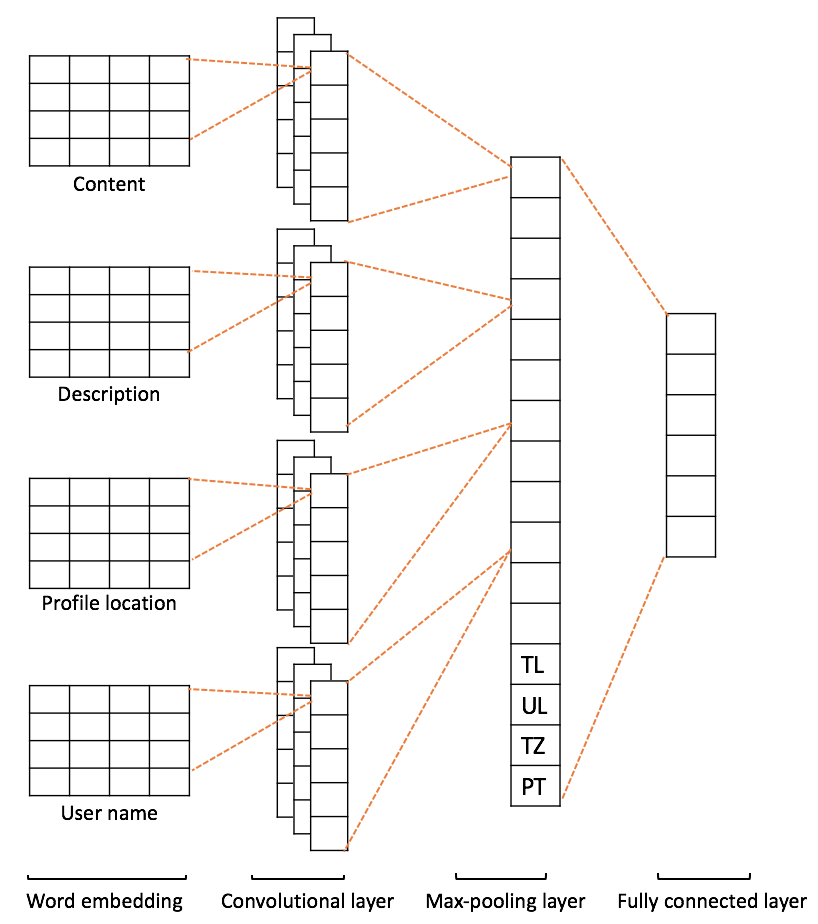}
  \vspace{-0.2cm}
    \caption{A diagram of the architecture of our neural model.}
      \vspace{-0.4cm}
    \label{arch}
\end{figure}

\vspace{-0.1cm}
Our approach is based on the convolutional neural network for sentence classification proposed by Kim\cite{kim2014convolutional}. Different from traditional bag-of-words method, such convolutional neural networks take the word order into consideration. Our model architecture is shown in Figure \ref{arch}. We use this CNN architecture to extract high-level features from four text fields in a tweet. Let $x_i^t\in R^k$ be the k-dimensional word vector corresponding to the $i$-th word in the text $t$, where $t\in$ \{tweet content, user description, profile location, user name\}. As a result, one text of length $n$ can be represented as a matrix \begin{equation}
X^t_{1:n} = x_1^t \oplus x_2^t \oplus ....\oplus x_n^t
\end{equation} where $\oplus$ is concatenation operator. In the convolutional layer, we apply each filter $w \in R^{hk}$ to all the word vector matrices, where $h$ is the window size and $k$ is the length of a word vector. For example, applying filter $w$ to a window of word vectors $x_{i:{i+h-1}}^t$, we generated $c_i^t = f(w\cdot x_{i:{i+h-1}}^t+b)$. Here $b\in R$ is a bias term and we choose $f(x)$ as a non-linear ReLU function $max(x,0)$. Sliding the filter window from the beginning of a word matrix till the end, we generated a feature vector $c^t=[c_1^t,c_2^t,...,c_{n-h+1}^t]$ for each text $t$. If we have $m$ filters in the convolutional layer, then we can produce $m$ feature vectors for each text field and $4m$ vectors in total. 

In the max-pooling layer, we apply a pooling operation over each feature vector generated in the convolutional layer. Each pooling operation takes a feature vector as input and outputs the maximum value $\hat c^t = max(c^t)$. $\hat c^t$ can be viewed as the most representative feature generated by a filter on text $t$. Hence we finally got a long vector $\theta \in R^{4m}$ after the max-pooling layer.  To avoid the co-adaptation of hidden units, we apply dropout on the max-pooling layer that randomly set elements in $\theta$ to zero in the training phase. After that, we append four categorical features tweet language(TL), user language(UL), timezone(TZ) and posting time(PT) with one-hot encoding at the end of $\theta$ and get $\hat \theta$. In the last fully connected layer, we use a softmax function over this long vector $\hat \theta$  to generate the probability distribution over locations. Specifically, the probability of one tweet coming from location $l_i$ is
 \begin{equation} P(l_i|\hat \theta) = \frac{exp(\beta_i^T\hat \theta)}{\sum_{j=1}^Lexp( \beta_j^T\hat \theta)}\label{prob}\end{equation} 
 where $L$ is the number of locations and $\beta_i$ are parameters in softmax layer. The output predicted location is just the location with highest probability. 

The minimization objective in the training phase is the categorical cross-entropy loss. The parameters to be estimated include word vectors, weight vectors $w$ for each filters, the weight vectors $\beta$ in softmax layer, and all the bias terms. The optimization is performed using mini-batch stochastic gradient descent and back-propagation\cite{rumelhart1988learning}.

%% file: 4_data.tex
\section{Data}
\vspace{-0.2cm}
We used geo-tagged tweets collected from Twitter streaming API\footnote{https://dev.twitter.com/streaming/reference/post/statuses/filter} for training and evaluation. In this study, we set the geographic bounding box as [-180, -90, 180, 90] so that we could get these geo-tagged tweets from the whole world. Our collection started from January 7, 2017 to February 1, 2017. Because it is very common for one user to post tweets from the same city, we randomly chose one tweet for each city that one user has visited. This could ensure that there is no strong overlap among our data samples. We only used tweets either with specific geo-coordinates or a geo-bounding box smaller than [0.1,0.1]. For the latter case, we used the center of one tweet's bounding box as its coordinates. No other filtering was done. There are 3,321,194 users and 4,645,692 tweets in total. For test data we used all tweets from 10\% of the users who were randomly selected. For the remaining 90\% users, we picked tweets from 50,000 of them as a development set and used the remaining tweets as training data.

There are two location prediction tasks we consider in this paper. The first task is country-level location prediction. We adopted the country code in the geo-tagged tweet as the label we want to predict. In our dataset, there are 243 countries and regions in total. The second task is city-level location classification. We adopt the same city-based representation as Han et al.\cite{bo2012geolocation}. The city-based representation consists of 3,709 cities throughout the world and was obtained by aggregating smaller cities with the largest nearby city. We assigned the closest city for each tweet based on orthodromic distance. Table \ref{data} contains basic statistics about our dataset. It is worth mentioning that this dataset is rather imbalanced, where a majority of tweets are sent from a few countries/cities. 
\begin{table*}[!h]
\centering
\vspace{-0.3cm}
\caption{Summaries about the dataset. Numbers in brackets are standard deviation.}
\label{data}
\begin{tabular}{c|c|c|c|c|c|c|c}
\hline
 \begin{tabular}[c]{@{}l@{}}\# of\\ tweets\end{tabular} & \begin{tabular}[c]{@{}l@{}}\# of\\ users\end{tabular} & \begin{tabular}[c]{@{}l@{}}\# of\\ timezones\end{tabular} & \begin{tabular}[c]{@{}l@{}}\# of\\ lang.\end{tabular} & \begin{tabular}[c]{@{}l@{}}\# of countries\\ (or regions)\end{tabular} & \begin{tabular}[c]{@{}l@{}}Tweets per\\country\end{tabular} &  \begin{tabular}[c]{@{}l@{}}\# of\\cities\end{tabular}&  \begin{tabular}[c]{@{}l@{}}Tweets per\\ city\end{tabular} \\ \hline
4645692      & 3321194        & 417             & 103             & 243                                                                    & 19118.0 (99697.1)      & 3709         & 1252.5(4184.5)       \\ \hline
\end{tabular}
\vspace{-0.5cm}
\end{table*}

%% file: 5_exp_v2.tex
\section{Experiments}
\vspace{-0.2cm}
\subsection{Evaluation Measures}
\vspace{-0.2cm}
Following previous works of tweet geolocation prediction\cite{han2013stacking}, we used four evaluation measures listed below. One thing to note is that when we calculated the error distance we used distance between predicted city and the true coordinates in the tweet rather than the center of assigned closest city.
\begin{itemize}
\item Acc: The percentage of correct location predictions.
\item Acc@Top5: The percentage of true location in our top 5 predictions.
\item Acc@161: The percentage of predicted city which are within a 161km(100 mile) radius of the true coordinates in the original tweet to capture near-misses. This measure is only tested on city-level prediction.
\item Median: The median distance from the predicted city to the true coordinates in the original tweet. This measure is only tested on city-level prediction.
\end{itemize}

\subsection{Baseline Method}
\vspace{-0.2cm}
We compared our approach with one commonly used ensemble method in previous research works \cite{han2013stacking,han2014text}. We implemented an ensemble classifier based on stacking\cite{wolpert1992stacked} with 5-fold cross validation. The training of stacking consists of two steps. First, five multinomial naive Bayes base classifiers are trained on different types of data(tweet content, user description, profile location, user name and the remaining categorical features). The outputs from the base classifiers are used to train a multinomial naive Bayes classifier in the second layer. We call such method STACKING in this paper. Same as \cite{han2014text}, we also use information gain ratio to do feature selection on text tokens. We call STACKING with feature selection STACKING+.

\subsection{Hyperparameters and Training}
\vspace{-0.2cm}
We used a tweet-specific tokenizer provided by NLTK\footnote{http://www.nltk.org/api/nltk.tokenize.html} to tokenize text fields. We built our dictionary based on the words that appeared in text, user description, and profile location. To reduce low-utility words and noise, we removed all words that had a word frequency less than 10. For our proposed approach, we used filter windows(h) of 3,4,5 with 128 feature vectors each, a dropout rate of 0.5 and batch size of 1024. We initialize word vectors using word2vec\footnote{https://code.google.com/archive/p/word2vec/} vectors trained on 100 billion words from Google News. The vectors have dimensionality of 300 and were trained using the continuous bag-of-words architecture\cite{mikolov2013distributed}. For those words that are not included in word2vec, we initialized them randomly. We also performed early stopping based on the accuracy over the development set. Training was done through stochastic gradient descent using Adam update rule with learning rate $10^{-3}$\cite{kingma2014adam}. For our baseline models, we applied additive smoothing with $\alpha=10^{-2}$, which is selected on the development set. For STACKING+ method, we first ranked these words by their information gain ratio value, then selected the top n\% words as our vocabulary. The tuning of n is based on accuracy over the development set. We selected n as 40\%, 55\% for city-level prediction and country-level prediction respectively.

\subsection{Results}
\vspace{-0.2cm}
The comparison results between our approach and the baseline methods are listed in Table \ref{country-table} and Table \ref{city-table}. Our approach achieves  92.1\% accuracy and 52.8\% accuracy on country-level and city-level location prediction respectively.  Our approach is consistently better than the previous model on the country-level location prediction task as shown in Table \ref{country-table}. It greatly outperforms our baseline methods over all the measures, especially on the city-level prediction task. It could assign more than half of the test tweets to the correct city and gain more than 20\% relative improvement over the accuracy of the STACKING+ method.
\vspace{-0.4cm}

\begin{figure}[!h]
         \vspace{-0.5cm}
     \includegraphics[width=0.5\textwidth]{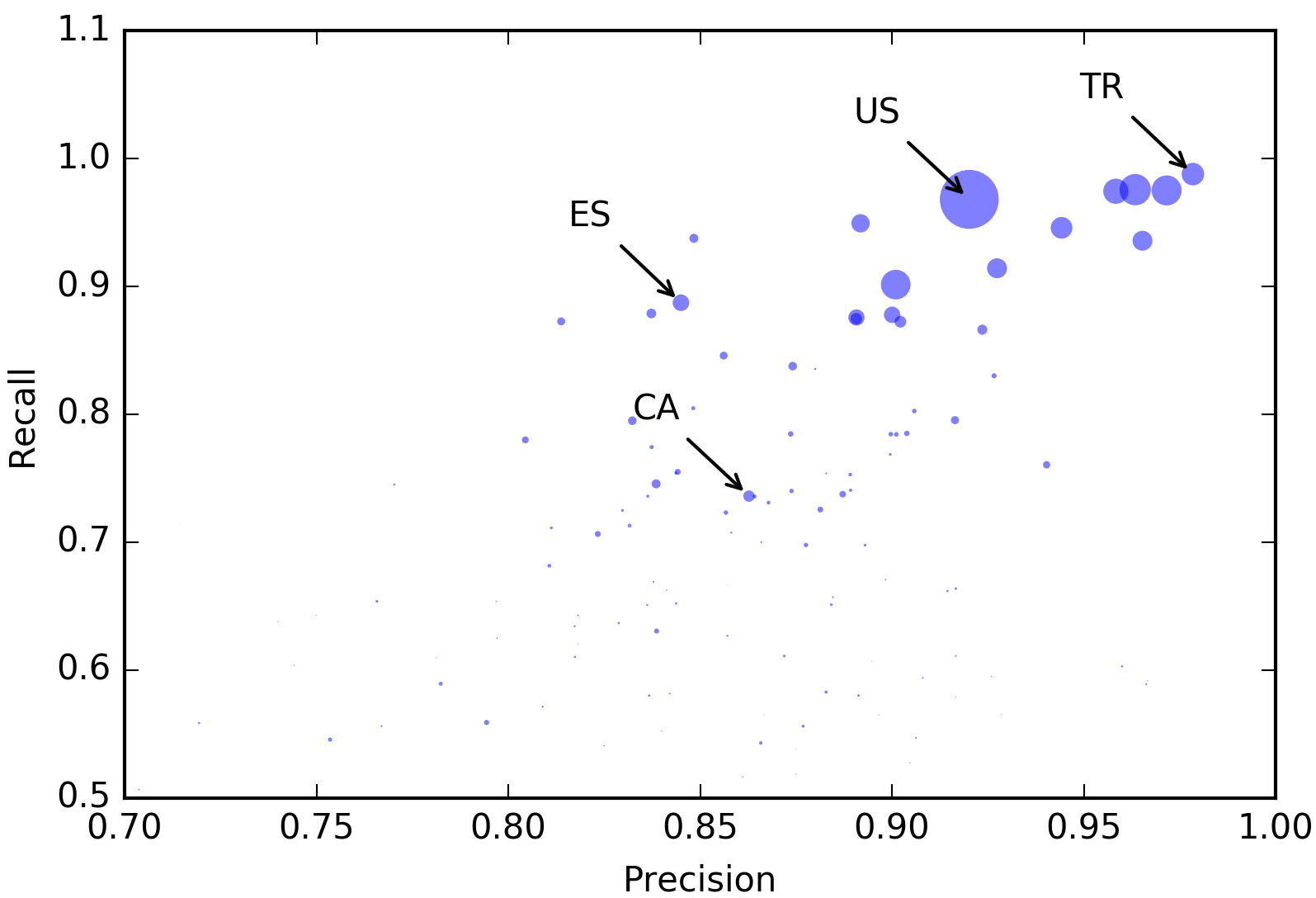}
         \includegraphics[width=0.5\textwidth]{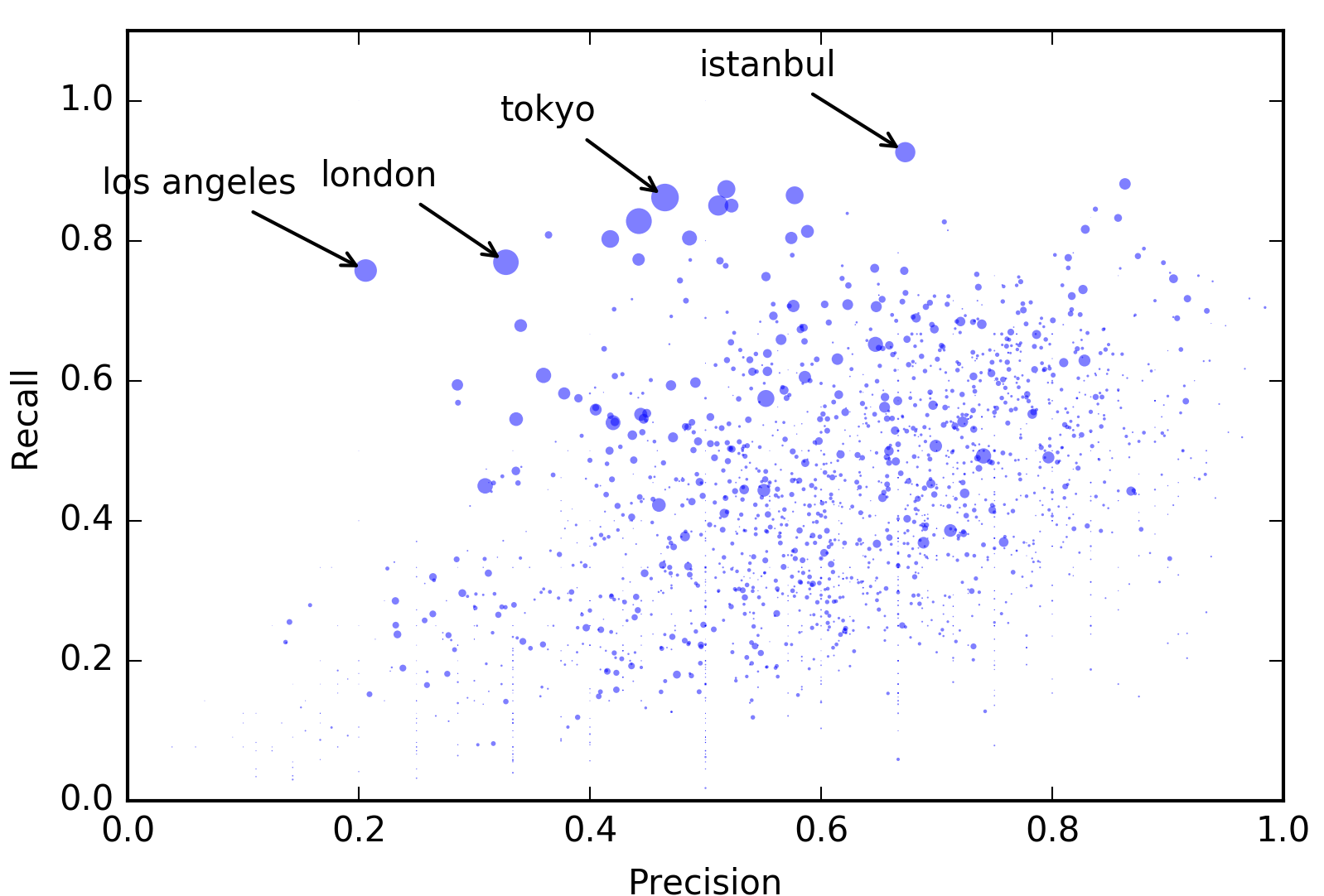}
         \vspace{-0.5cm}
     	        \caption{Two scatter graphs that show the performances for each country and cities. The x-axis is precision, y-axis is recall. Each dot represents a country/city. The dot size is proportional to the number of tweets that comes from the correponding location. Some tiny invisible country outside of the scope are not shown in the figure.}
     	                 \vspace{-0.2cm}
    \label{dot_country}
    	\label{dot_city} 
\end{figure}

\begin{table}
\parbox{.42\linewidth}{
\centering
\caption{Country prediction results.}
\label{country-table}
\begin{tabular}{|l|l|l|}
\hline
             & Acc   & Acc@Top5 \\ \hline
STACKING     & 0.868 & 0.947   \\ \hline
STACKING+    & 0.871 & 0.950   \\ \hline
Our approach & \textbf{0.921} & \textbf{0.972}   \\ \hline
\end{tabular}
}
\hfill
\parbox{.58\linewidth}{
\centering
\caption{City prediction results.}
\label{city-table}
\begin{tabular}{|l|l|l|l|l|}
\hline
             & Acc   & Acc@161 & Acc@Top5 & Median \\ \hline
STACKING     & 0.389 & 0.573   & 0.595    & 77.5 km  \\ \hline
STACKING+    & 0.439 & 0.616   & 0.629    & 47.2 km  \\ \hline
Our approach &\textbf{0.528} & \textbf{0.692}   & \textbf{0.711}    & \textbf{28.0} km   \\ \hline
\end{tabular}
}
\end{table}
\vspace{-0.4cm}
Our approach performs better for countries with a large number of tweets. In Figure \ref{dot_country}, we plotted the precision and recall value for each country as a scatter chart. The dot size is proportional to the number of tweets that come from that country. Turkey appears to be the country with highest precision and recall. These results suggest that our approach works better with more data samples.

The same graph is also plotted for city prediction in Figure \ref{dot_city}. Because of the skewness of our data and the difficulty of city-level prediction, our classifier tends to generate labels towards big cities, which leads to high recall and low precision for cities like Los Angeles.

\begin{figure}[h]
\vspace{-0.3cm}
	     \includegraphics[width=0.5\textwidth]{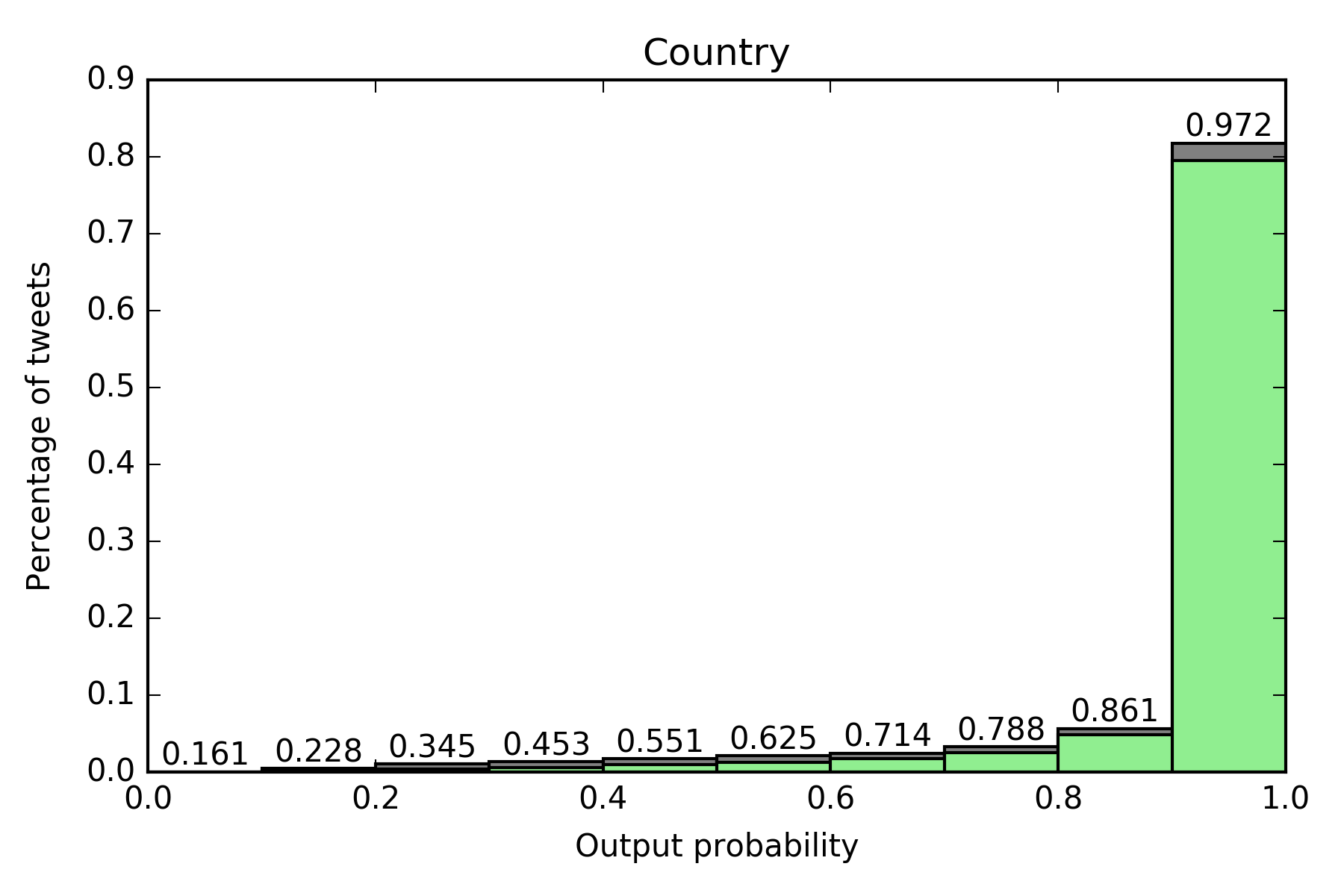}
         \includegraphics[width=0.5\textwidth]{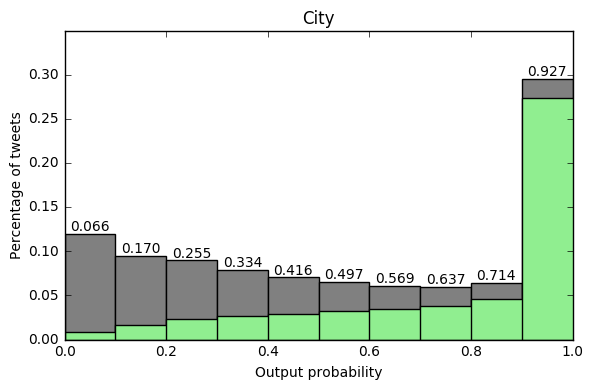}
              	        \vspace{-0.5cm}
    \caption{Two bar charts that show the distribution of tweets in terms of the output probability. The x-axis is the output probability associated with each prediction, the y-axis is the percentage of tweets. The height of grey bar represents the percentage of test data that has certain prediction output probability. The height of green bar represents the percentage of correctly predicted tweets in each output probability range. We listed the accuracy for each probability range above the bar. Take the rightmost bar in country-level prediction for example, there are 81.8\% test tweets' country are predicted with output probability larger than 0.9. Among these 81.8\% tweets, 97.2\% are predicted correctly.}
         	        \vspace{-1cm}
    	\label{dist_city}   
\end{figure}

For real world applications, people may ask how we could set a threshold to get prediction results with high confidence. To answer this question, we further examined the relation between prediction accuracy and the output probability. Here the output probability is just the probability of our predicted location calculated by equation \ref{prob}. Figure \ref{dist_city} shows the distribution of tweets in terms of output probability for two tasks. The grey bar represents the percentage of testing tweets within certain range of output probability. The green bar represents the percentage of correctly classified tweets. The number on the grey bar is the accuracy for each output probability range. 

As expected, the prediction accuracy increases as the output probability increases. We  get 97.2\% accuracy for country-level prediction  with output probability larger than 0.9. Surprisingly, the accuracy of city-level is as high as 92.7\% for the 29.6\% of the tweets with output probability greater than 0.9. However, the city-level accuracy for the remaining tweets with output probability less than 0.9 is only 48.4\%. Unlike country-level prediction, the number of tweets decreases as output probability increases, unless the output probability is larger than 0.9. 

%% file: 6_conclusion.tex
\section{Discussion and Conclusion}
     	        \vspace{-0.2cm}
These experiments demonstrate that our approach is consistently better than the prior method thus supporting more tweets to be accurately located by country, and city, of origin. At the country level, the more tweets that come from the country, the better the prediction. Regardless of the number of tweets per country, we can predict the country location for most tweets wih extremely high confidence and accuracy. At the city level the results are more mixed. For a small fraction of tweets we can get greater than 90\% accuracy, but for the rest of tweets the accuracy is less than 50\%. For about half the tweets it is difficult to infer the city location. This result is partially due to the fact that we base the prediction on only a single tweet. Future work may consider using collection of tweets per user. This result is also partially due to the fact that the data is highly skewed toward a few cities. Future work should develop a training set that is more evenly distributed across cities. Despite these limitations, this approach shows promise.

This paper presents a method for geo-locating a single tweet based on the information in a tweet JSON object. The proposed approach integrates tweet text and user profile meta-data into a single model. Compared to the previous stacking method with feature selection, our approach substantially outperforms the baseline method.  We developed the approach for both city and country level and demonstrated the ability to classify tweets at both levels of granularity. The results demonstrate that using a convolutional neural network utilizes the textual location information better than previous approaches and boosts the location prediction performance substantially.